\documentclass[aps,twocolumn,superscriptaddress,showpacs,groupedaddress,reprint]{revtex4}  % for review and submission

\usepackage{graphicx}  % needed for figures
\usepackage{dcolumn}   % needed for some tables
\usepackage{bm}        % for math
\usepackage{amssymb}   % for math
\usepackage{xspace}
\usepackage{amsmath}
\usepackage{abbrevs}
\usepackage[]{units}	%Auto-formats units
\usepackage{color}

\usepackage{float} 
\usepackage{wrapfig} 
\usepackage{multirow} % Required for multirows
\usepackage{rotating} %rotating tables

\begin{document}

\title{ Ultratunable quantum frequency conversion in photonic crystal fiber}

\author{K.~A.~G. Bonsma-Fisher}
\affiliation{National Research Council of Canada, 100 Sussex Drive, Ottawa, Ontario K1A 0R6, Canada}
\author{P.~J. Bustard}
\email[]{philip.bustard@nrc-cnrc.gc.ca}
\affiliation{National Research Council of Canada, 100 Sussex Drive, Ottawa, Ontario K1A 0R6, Canada}
\author{C. Parry}
\affiliation{Centre for Photonics and Photonic Materials, Department of Physics, University of Bath,  Bath, BA2 7AY, UK}
\author{T.~A. Wright}
\affiliation{Centre for Photonics and Photonic Materials, Department of Physics, University of Bath,  Bath, BA2 7AY, UK}
\author{D.~G. England}
\affiliation{National Research Council of Canada, 100 Sussex Drive, Ottawa, Ontario K1A 0R6, Canada}
\author{B.~J. Sussman}
\affiliation{National Research Council of Canada, 100 Sussex Drive, Ottawa, Ontario K1A 0R6, Canada}
\affiliation{Department of Physics, University of Ottawa, Advanced Research Complex, 25 Templeton Street, Ottawa, Ontario K1N 6N5, Canada}
\author{P.~J. Mosley}
\email[]{p.mosley@bath.ac.uk}
\affiliation{Centre for Photonics and Photonic Materials, Department of Physics, University of Bath,  Bath, BA2 7AY, UK}

\date{\today}

\begin{abstract}
\noindent Quantum frequency conversion of single photons between wavelength bands is a key enabler to realizing widespread quantum networks.
We demonstrate the quantum frequency conversion of a heralded 1551\,nm photon to any wavelength within an ultrabroad (1226 - 1408\,nm) range in a group-velocity-symmetric photonic crystal fiber (PCF), covering over 150 independent frequency bins. The target wavelength is controlled by tuning only a single pump laser wavelength. 
We find internal, and total, conversion efficiencies of 12(1)\% and 1.4(2)\%, respectively. 
For the case of converting 1551\,nm to 1300\,nm we measure a heralded $g^{(2)}(0) = 0.25(6)$ for converted light from an input with $g^{(2)}(0) = 0.034(8)$.
We expect that this PCF can be used for a myriad of quantum networking tasks.
\end{abstract}

\pacs{42.50.-p, 42.65.Ky, 42.81.-i}

\maketitle

%%% FIG %%%
\begin{figure*}[!t]
%\center{\includegraphics[width=1.5\columnwidth]{FIG1.pdf}}
%\center{\includegraphics[width=2\columnwidth]{FIG1_row.pdf}}
\center{\includegraphics[width=2.\columnwidth]{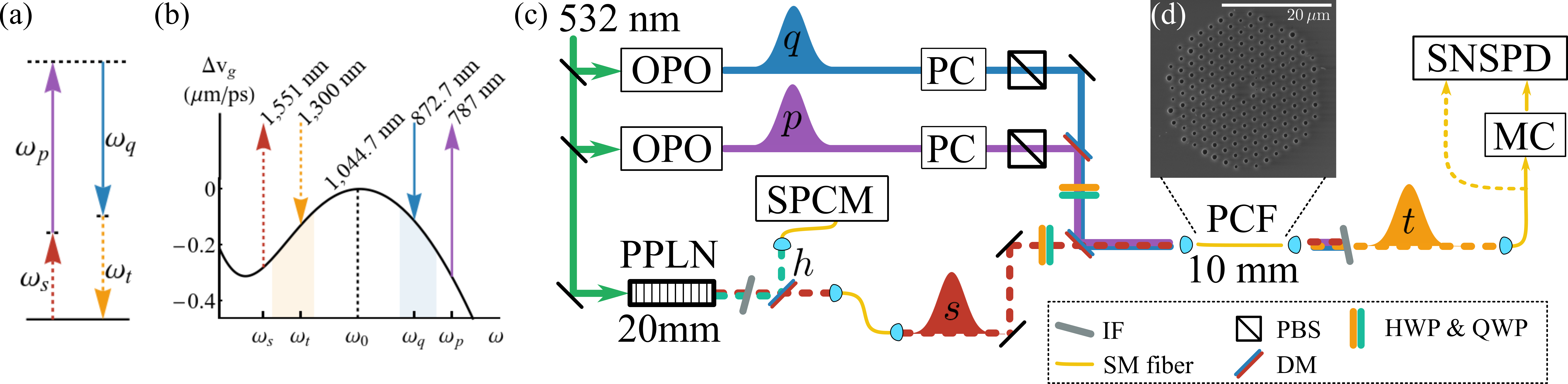}}
\caption{
(a) Four-wave mixing process leading to frequency conversion from an input signal at $\omega_s$ to a target at $\omega_t$.
(b) The PCF group velocity remains symmetric about the zero dispersion frequency ($\omega_0$) across an ultrabroad range. Signal and pump fields are group velocity matched on opposite sides of the zero dispersion point; $\omega_p$ is matched to $\omega_s$, and $\omega_q$ to $\omega_t$. By changing $\omega_q$ (blue shading), the target frequency $\omega_t$ is tuned across an ultrabroad range (orange shading). 
(c) Experimental setup. Pump fields $p$ and $q$, as well as input photon $s$ (see main text for details), are coupled into the PCF. Upon exiting the PCF, converted photon $t$ is collected in to SM fiber and to SNSPDs, or first to a scanning monochromator (MC).  Half-(quarter-)wave plate (HWP, QWP); interference filter (IF); polarizing beam splitter (PBS); dichroic mirror (DM). 
(d) Scanning electron micrograph of the PCF with a pitch (inter-hole spacing) of $\Lambda=2.11$\,$\mu$m and hole diameter $d=0.71$\,$\mu$m.
} \label{fig:setup}
\end{figure*}

\noindent The Quantum Internet~\cite{Kimble2008} is poised to connect users spanning a wide array of quantum capabilities and needs. 
Quantum systems are now being entangled across large distances in metropolitan environments~\cite{Yu2020}.
While single photons are the natural choice to carry quantum information between network components, there is no single wavelength which suits all quantum networking needs.
The essential elements of a quantum network: quantum processors~\cite{Politi2008}, quantum memories~\cite{Kozhekin2000}, photon sources~\cite{Kurtsiefer2000}, and detectors~\cite{Lita2008}, whose implementations cover a broad range of physical systems, can have vastly different spectro-temporal requirements for coupling quantum light.

Quantum frequency conversion (QFC)~\cite{Kumar1990} is a powerful tool to hybridize diverse quantum nodes that operate at different wavelengths and away from the low-loss bands of telecom fiber networks~\cite{Kielpinski2011}.
A striking example~\cite{Maring2017} is the coherent networking of a cold atomic ensemble (780\,nm) to a rare-earth doped crystal (606\,nm) via a telecom link (1552\,nm).
Moreover, QFC is also a workhorse for quantum information processing~\cite{Langford2011, Eckstein2011, Humphreys2014, Joshi2022}, photon source multiplexing~\cite{Joshi2018}, quantum memory~\cite{Bustard2022}, and ultrafast measurement~\cite{Donohue2014, Maclean2018}.
QFC has been demonstrated in numerous systems spanning bulk crystals~\cite{Huang1992, Fisher2016}, crystal waveguides~\cite{Kasture2016,Bock2018,Walker2018}, silicon micro-resonators~\cite{Li2016}, molecular gases~\cite{Bustard2017, Tyumenev2022}, and fiber optics~\cite{McGuinness2010, Clark2013}.
QFC demonstrations using second-order ($\chi^{(2)}$) nonlinearities, in periodically-poled lithium niobate (PPLN) waveguides~\cite{Maring2018} for instance, can display internal conversion efficiencies of over 90\%.
There, a single pump field mediates the conversion, either by sum- or difference-frequency generation, of an input signal to its target wavelength.
This strategy is often used for large frequency shifts, linking telecom and optical frequencies~\cite{Albrecht2014}, though small shifts are possible by cascading processes~\cite{Fisher2021}.
While these processes can be made efficient, phase matching is usually achieved in a narrow window making the conversion efficiency highly sensitive to the crystal poling period and pump wavelength, unless advanced poling patterns are used to access the adiabatic conversion regime~\cite{Suchowski2009}.
Practically, this means each use case requires a new device with a precisely chosen poling period.

Third-order nonlinear ($\chi^{(3)}$) processes use two pump fields which can offer robust control of the target photon's frequency and spectro-temporal profile~\cite{Mckinstrie2005}.
Bragg scattering four-wave mixing (BSFWM) is a particularly attractive option since it is inherently low-noise, can be used for both small and large frequency shifts including inter- and intra-band telecom conversion, and can be implemented using off-the-shelf fiber optic components~\cite{Christensen2018}.
In BSFWM, two pump fields ($p$ and $q$) convert an input signal $s$ to a target $t$, such that $\omega_t = \omega_s + (\omega _p - \omega_q)$, following energy conservation (see Fig.~\ref{fig:setup}(a)).
Phase matching is achieved when the four fields have equal and opposite pairwise detunings from the zero dispersion frequency $\omega_0$, or by using birefringence in a polarization-maintaining (PM) fiber. 
In practice, with standard or PM fiber, efficient BSFWM occurs only for specific combinations of pump, source, and target wavelengths for which the fiber dispersion enables phase matching to be satisfied.
Changes in source or target wavelength require at least the tuning of both pump wavelengths and, often, the selection of a new fiber with different dispersion. 

Photonic crystal fibers (PCF) are a leading platform for BSFWM~\cite{McGuinness2010,Wright2020} since their structure can be engineered to control the group velocity dispersion while maintaining low-loss guidance in the fundamental mode.
In this work we demonstrate a QFC device that can span a wide array of use cases in a quantum network, maintaining efficient BSFWM in a single PCF across an ultrabroad wavelength range. 
To achieve this, we have fabricated a PCF that has a group velocity profile which remains symmetric about the zero dispersion frequency for over 1\,PHz~\cite{Parry2021}.
This gives an ultrabroad phase matching window, meaning that the pump frequency $\omega_p$ can remain fixed opposite the input signal $\omega_s$, while the second pump frequency $\omega_q$, can be tuned to determine the target frequency $\omega_t$.
We observe QFC of an input photon at $\lambda_s = 2\pi c/\omega_s =1551$\,nm to target wavelengths across the range $\lambda_t = 1226-1482$\,nm, spanning the entire telecom o- and e-bands.
This PCF allows a single device to be used for numerous spectro-temporal requirements, where a single pump wavelength is tuned to tailor its use to a given quantum network node. 
The bidirectional nature of BSFWM enables the same device to convert source photons anywhere in the range 1226-1482\,nm to the same target wavelength, 1551\,nm.

The PCF in this experiment is fabricated from ultra-pure silica using the stack-and-draw technique. 
The cladding consists of a triangular array of air holes with nominal pitch $\Lambda = 2.11$\,$\mu$m and hole diameter to pitch ratio $d/\Lambda=0.337$.
A solid core is formed by omitting one hole from the array as shown in the micrograph of the cleaved end face in  Fig.~\ref{fig:setup}(d).
The fiber is designed for a zero dispersion frequency of $\omega_0 = 2\pi c/ (1044.7\text{\,nm}) $ and guides single modes across all wavelengths in this experiment.
The group velocity profile of the PCF, calculated from equations in Ref.~\cite{Saitoh2005}, is shown in Fig.~\ref{fig:setup}(b). 
The output signal photon frequency $\omega_t$ can be tuned across the shaded frequency range by tuning $\omega_q$ across the complementary range on the opposite side of the zero dispersion frequency.
We use a 10\,cm PCF, combined with 0.1\,nm pump bandwidths, to give a broad tuning range. 
For this length, $p$ and $q$ pulses walkoff from each other by no more than 0.6\,ps due to chromatic dispersion.
%Longer lengths of PCF could then be used to increase conversion efficiency, however the range over which conversion is phase-matched for a single $\lambda_p$ value scales as $1/L^2$. 
%This range can be recovered by using broader pump bandwidths: for 5\,nm pump bandwidths the tuning range remains for metre-scale PCF lengths.

The experimental setup is shown in Fig.~\ref{fig:setup}(c). The primary laser for this experiment (Coherent Paladin) outputs $\sim 15$\,ps pulses with wavelength 532\,nm at an 80\,MHz repetition rate. The 532\,nm beam is split to pump two optical parametric oscillators (OPO, APE Levante Emerald), producing $p$ and $q$ fields with measured pulse durations 13.4\,ps and 14.7\,ps, respectively. The pump wavelength $\lambda_p$ is set to 787.0\,nm, unless otherwise stated, while $\lambda_q$ is tuned across the range 810-910\,nm. 
A 532\,nm pulse also pumps spontaneous parametric downconversion (SPDC) in a periodically-poled lithium niobate (PPLN) crystal to produce signal ($s$) and herald ($h$) photon pairs near 1550\,nm and 810\,nm, respectively. 
The $h$ photon is collected in a single-mode (SM) fiber and its detection on a single-photon counting module (SPCM) heralds the presence of an $s$ photon directed to the PCF. 
The coupling efficiency of $s$ into the PCF is $\eta^\text{in}_s = 0.44(4)$.
Due to a long-lived fluorescence noise process in the PCF, we pick pump pulses $p$ and $q$ at a rate of 4.71\,MHz using a Pockels cell (PC).
Pumps are then combined on a dichroic mirror and coupled into the PCF with 48\% and 54\% efficiencies, respectively, with typical pulse energies of 6\,nJ in the PCF. 
The polarizations, pulse arrival times, and pulse energies of $p$ and $q$ are independently controlled.
Upon exiting the PCF, $p$ and $q$ are blocked by frequency filters, $s$ and $t$ are split by a 1500\,nm short-pass dichroic mirror and are coupled into SM fibers.

%%% FIG %%%
\begin{figure}
\center{\includegraphics[width=1.\columnwidth]{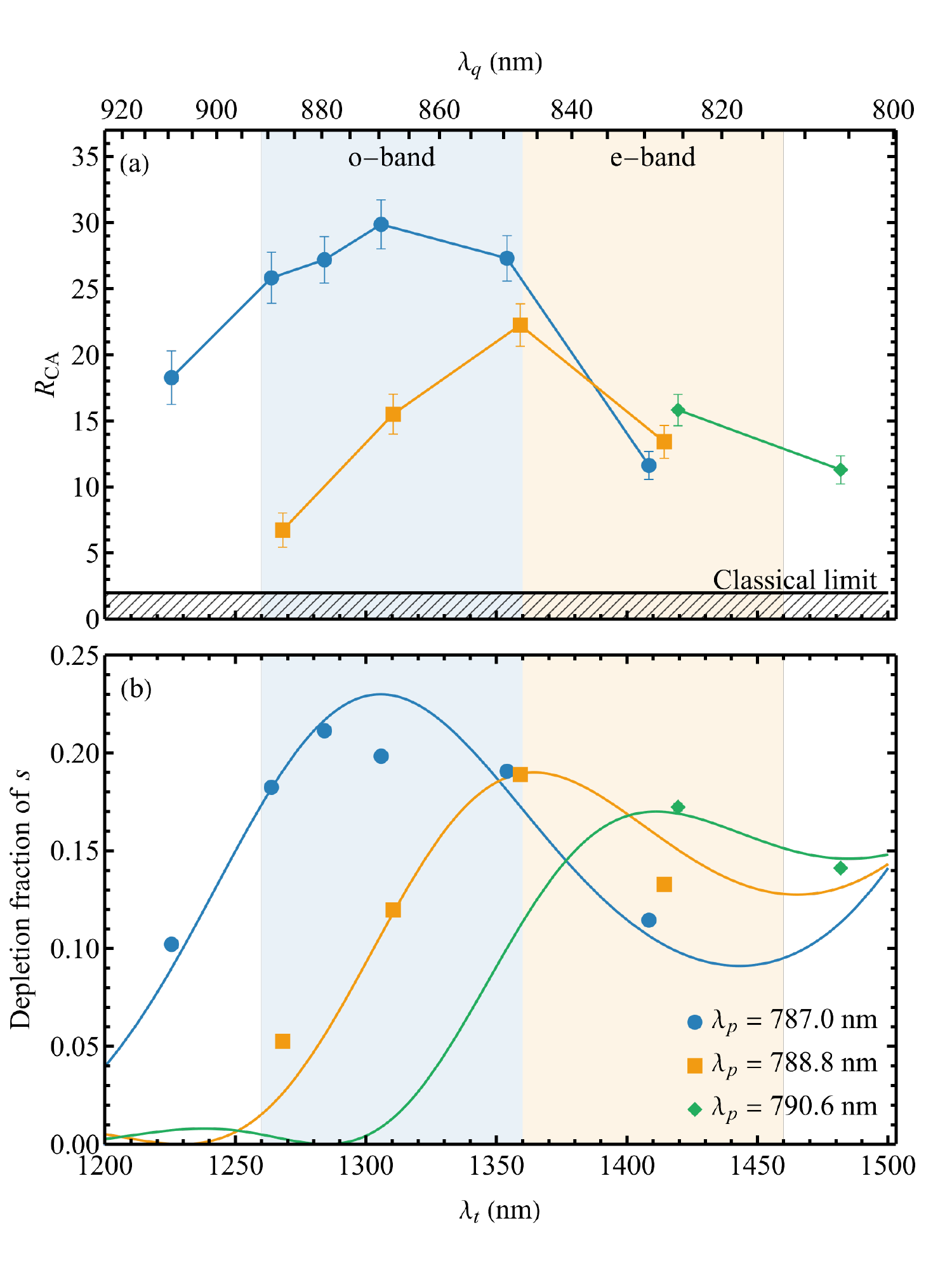}}
\caption{ (a) Coincidence-to-accidental ratio $R_{CA}$ between $t$ and $h$ photons is $\gg 1$ across an ultrabroad range of frequency conversion, including the entire telecoms o- and e-bands (blue and orange shaded regions, respectively). Uncertainties are derived from Poissonian error in photon detection. 
(b) Fraction of 1551\,nm source light depleted by pumps. Solid lines in (b) show theoretical conversion efficiencies rescaled for comparison with the data. 
} \label{fig:g2xc}
\end{figure}

%%% FIG %%%
\begin{figure}[h!]
\center{\includegraphics[width=1.\columnwidth]{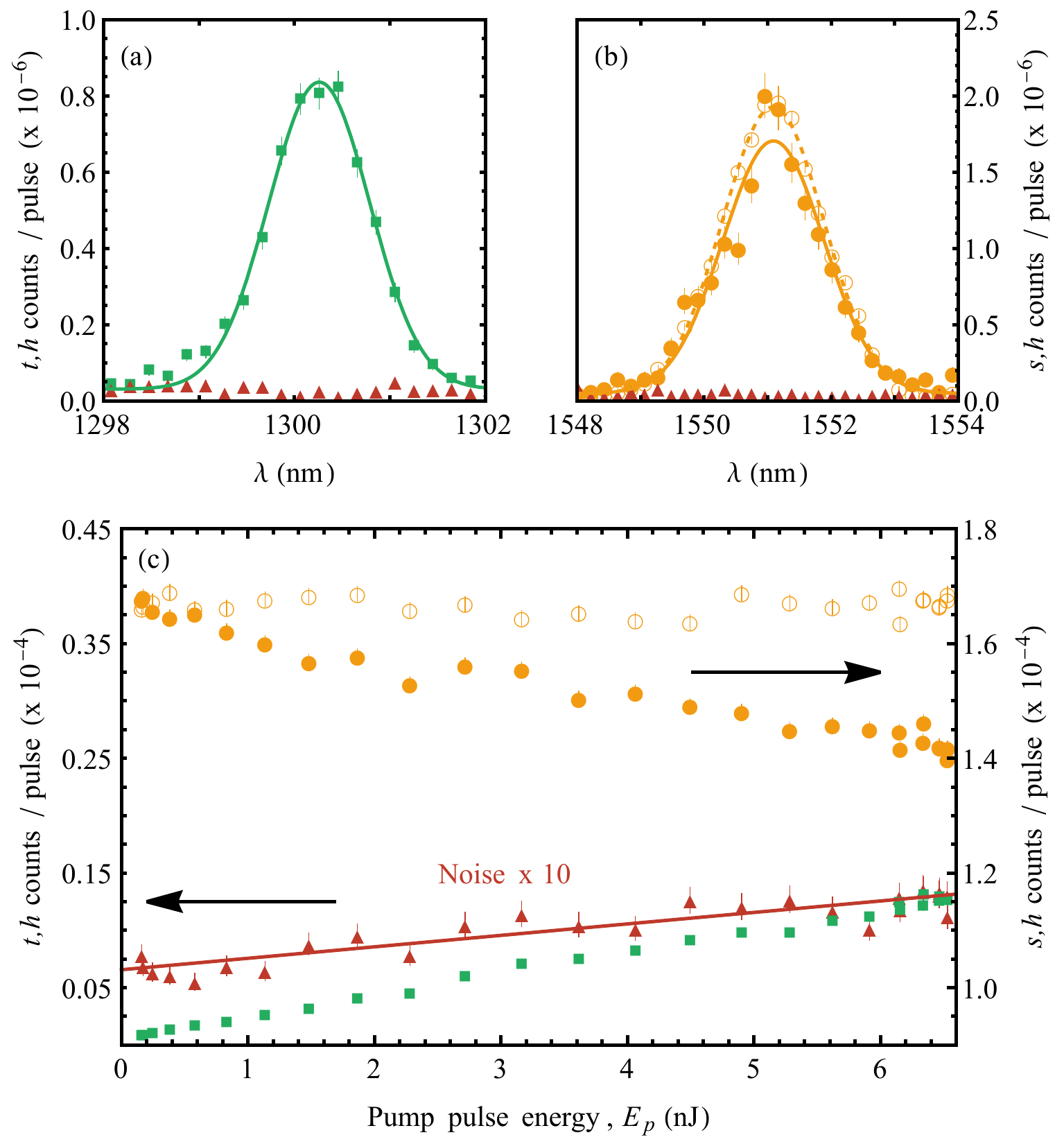}}
\caption{ Monochromator (a,b), and pump power (c) scans when $\lambda_p=787.0$\,nm, $\lambda_q=872.7$\,nm. 
In (a-c), coincidence counts are between herald ($h$) and converted ($t$) photons (left axis, green squares), noise photons (left axis, red triangles), input ($s$) photons with(out) pumps (right axis, solid (hollow) orange circles). Error bars are derived from Poissonian error in photon detection. Solid and dashed lines show Gaussian fits to data.
(a) Converted spectrum at 1300\,nm.
(b) Input photons, with no pumps present (hollow orange circles) and  with pumps present (filled orange circles).
(c) Scan of $p$-pump pulse energy, with $q$ pulse energy fixed at 5.5\,nJ.  Noise counts are scaled by a factor of 10. Solid line is a linear fit to noise data. 
Noise counts have been subtracted from the depleted signal (filled orange circles, right axis) for direct comparison with the input.
}\label{fig:MCdlypow}
\end{figure}

To demonstrate the range of frequency conversion in the PCF, $t$ photons are sent to a scanning monochromator (MC) with 0.5\,nm resolution. 
Exiting the monochromator, photons are coupled into SM fiber, with total collection efficiency of 10\%, and measured on a superconducting nanowire single-photon detector (SNSPD). 
As $\lambda_q$ is scanned across the range 810-910\,nm, the monochromator is tuned to the optimal $\lambda_t$ setting and coincident detections between $t$ and $h$ photons are recorded using a 1\,ns coincidence window.
Unconverted $s$ photons are sent directly to a SNSPD and coincidences with $h$ detections are also recorded. 
In Fig.~\ref{fig:g2xc}(a) we plot the coincidence-to-accidental ratio, $R_{CA} = N_{t,h} \times N_\text{pulses} / (N_t N_h)$, where $N_i$ is the number of detections in the $i^\text{th}$ mode. 
Values of $R_{CA}>2$ are characteristic of nonclassical correlations between the signal and herald modes. 
At the input signal wavelength, we measure $R_{CA}=62.4(5)$.
With $\lambda_p$ fixed at 787\,nm, quantum frequency conversion is achieved for wavelengths 1226-1408\,nm. This spans the entire telecom o-band (1260-1360\,nm), with $R_{CA} > 25$.  
With small adjustments to $\lambda_p$ the tuning range can be further extended to longer wavelengths, encompassing the telecom e-band (1360-1460\,nm) as well. 
The results indicate that nonclassical correlations are retained through the BSFWM process.
In principle, the range can also be extended to shorter $\lambda_t$, however we observe a strong noise signal due to a seeded four-wave mixing process from the pumps when $\lambda_q \sim 920$\,nm; for even larger shifts we are limited by the OPO tuning range.
The exact conversion efficiency is difficult to calculate due to wavelength-dependent losses. 
Instead, Fig.~\ref{fig:g2xc}(b) shows by what fraction the input $s$ is depleted in the conversion process, placing an upper limit on conversion efficiency.
Shown alongside the data are scaled theoretical conversion efficiencies $\eta_\text{QFC} = \text{sinc}^2 \Delta \beta L/2$, where $L$ is the fiber length, and the phase mismatch $\Delta \beta = \beta_p + \beta_s - \beta_q - \beta_t$, with fundamental-mode propagation constants $\beta_i = n_\text{eff} (\omega_i) \omega_i / c$.
The effective index $n_\text{eff}(\omega)$ is fitted using well-known empirical relations for PCF~\cite{Saitoh2005} with $\Lambda$ and $d/\Lambda$ as free parameters. 
The best fit is achieved for $\Lambda = 2.1044$\,$\mu$m and $d/\Lambda=0.3389$, in close agreement with the nominal values.
We note that highest values of $R_{CA}$ are measured when the conversion efficiency is highest.

%%% TABLE %%%
\begin{table}
  \begin{center}
    \begin{tabular}{l|l}
      $g^{(2)}_{s,s}(0)$ (1551\,nm) & 0.034(8)\\
      $g^{(2)}_{t,t}(0)$ (1300\,nm) & 0.25(6) \\
      $g^{(2)}_{n,n}(0)$ (noise, 1300\,nm) & 0.97(4)
    \end{tabular}
  \end{center}
  \caption{ Measured $g^{(2)}_{i,i}(0)$ values for input ($i$), target ($t$, and noise ($n$) fields.   }
  \label{tab:g2auto}
\end{table}

To demonstrate the nonclassical nature of the target field, we fix the pump wavelengths at $\lambda_p = 787.0$\,nm and $\lambda_q=872.7$\,nm, converting $s$ photons at 1551\,nm to $\lambda_t = 1300$\,nm. 
Routing converted (Fig.~\ref{fig:MCdlypow}(a)) and unconverted (Fig.~\ref{fig:MCdlypow}(b)) light through the monochromator, we measure spectra for the $t$ and $s$ fields, respectively. 
The fitted bandwidths of $s$ and $t$ are 225\,GHz and 202\,GHz FWHM, after accounting for the MC resolution. 
Noise photons ($n$) --- measured when only pump beams are input to the PCF --- are relatively constant in the ranges of the $s$ and $t$ MC scans.

Next, we use tight in-line spectral filtering around the 1300\,nm and 1551\,nm beams after they are separated by the dichroic mirror, and route light directly to the SNSPDs with overall collection efficiencies of $\eta^\text{out}_t = 0.28(2)$ and $\eta^\text{out}_s = 0.54(5)$, respectively. 
Figure~\ref{fig:MCdlypow}(c) shows the dependencies of the converted (1300\,nm), and unconverted (1551\,nm), and noise fields on the $p$ pulse energy ($E_p$).
Accounting for the relative collection and detection efficiencies in the 1300\,nm and 1551\,nm paths, we estimate the internal conversion efficiency to be $\eta^\text{int}_\text{QFC}=0.12(1)$ at the highest pump pulse energy.
Including in- and out-coupling efficiencies between the PCF and SM fiber, the total conversion efficiency is $\eta_{QFC} = \eta^\text{in}_s \times \eta^\text{int}_\text{QFC} \times \eta^\text{out}_t = 0.014(2)$, similar to  total efficiencies shown in PPLN waveguides~\cite{Walker2018}.
We expect that the conversion efficiency can be improved in several ways. 
The conversion efficiency for BSFWM~\cite{McKinstrie2002, Lefrancois2015} in a medium of length $L$ is given by $\eta_\text{BSFWM} = (4\gamma^2 P_p P_q / \kappa^2) \sin^2 \kappa L$, where $\gamma$ is the nonlinear coefficient, $P_p$ and $P_q$ are the pump powers, and $\kappa = \sqrt{(\Delta \beta/2)^2 + 4\gamma^2 P_p P_q}$, assuming $P_p \sim P_q$. 
The observed near-linear scaling with $P_p$ in Fig.~\ref{fig:MCdlypow}(c) is consistent with this. 
Higher pump powers would further increase the efficiency while maintaining the current tuning range. 
Closer matching of the pump bandwidths to the signal would increase the tuning range and deliver higher pump powers at equivalent pulse energies. 
Alternatively, the fiber length could be increased; however, this reduces the frequency range over which conversion is phase-matched for a single $\lambda_p$ value, and group velocity walk-off will eventually limit the effective interaction length.

Next, we route $t$ photons to a two-element SNSPD, which allows us to distinguish between 0, 1 or 2 detections. 
We measure the second-order coherence function~\cite{HBT} of the converted light, input, and noise, calculated for a field $i$ as $g_{i,i}^{(2)}(0) = N_{1,2,h} \times N_\text{pulses} / (N_{1,h} N_{2,h})$, where the indices 1 and 2 refer to the outputs of the two-element detector.
A perfect single photon source has $g_{i,i}^{(2)}(0) = 0$, and any $g_{i,i}^{(2)}(0) < 1$ displays sub-Poissonian, i.e., nonclassical, statistics~\cite{MandelWolf}.
We recorded two- and three-fold coincidence data over a 40-hour span; the measured $g_{i,i}^{(2)}(0)$ values (Table~\ref{tab:g2auto}) confirm the quantum nature of the converted light.
%We recorded two- and three-fold coincidence data over a 40-hour span; and measured the values $g_{s,s}^{(2)}(0) = 0.034(8)$, $g_{t,t}^{(2)}(0) = 0.25(6)$, and $g_{n,n}^{(2)}(0) = 0.97(4)$, for the signal, target, and noise, respectively, confirming the quantum nature of the converted light.

%%% FIG %%%
\begin{figure}
\center{\includegraphics[width=0.87\columnwidth]{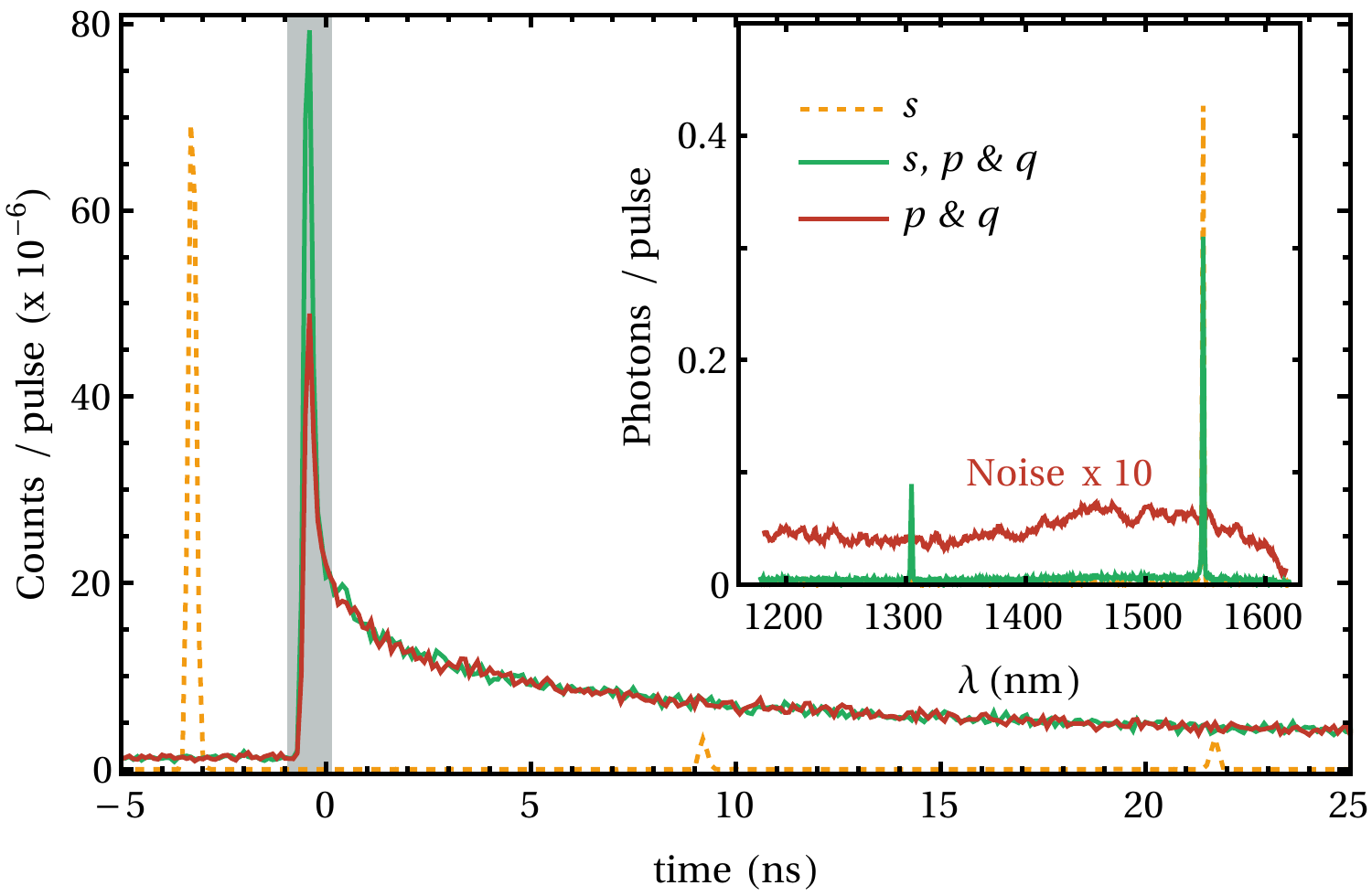}}
\caption{  
Detection times of 1550\,nm input photons (orange), 1300\,nm target (green) and noise (red) photons relative to the Pockels cell trigger.
The 1\,ns coincidence window is shown for reference (shaded gray).
\emph{Inset}: Measured spectrum showing broadband noise. Input and target spectra are shown for reference. Here, the input is attenuated 1550\,nm pulses with 1 photon/pulse on average.  
} \label{fig:noise}
\end{figure}

Finally, we discuss characteristics of the noise detected in this experiment. A histogram (Fig.~\ref{fig:noise}) of noise detection times, with respect to the pump pulse arrival, shows an exponential tail with $1/e$ lifetime of $10$\,ns, compared to the input photon detection peak and detector jitter (100\,ps).
The noise spectrum is relatively constant between 1200-1600\,nm (Fig.~\ref{fig:noise} inset); the quantum efficiency of the InGaAs detector drops sharply beyond 1600\,nm. 
This behaviour, and the linear scaling with pump pulse energy (Fig.~\ref{fig:MCdlypow}(c)), are consistent with fluorescence due to point defects in the silica.
The long histogram tail which dominates the noise can be filtered out by using a short, $\leq$1\,ns, coincidence window. 
We expect that the noise could be largely negated by treating the silica before, or after, fabrication to remove point defects without affecting the nonlinear properties of the PCF.

In summary, we have demonstrated QFC of heralded single photons from 1551\,nm to any wavelength in a $175$\,nm range in a single PCF while only tuning one pump wavelength (over $250$\,nm with small adjustments to $\lambda_p$).
The converted photons showed a coincidence-to-accidental ratio $R_{CA} >10$ for much of this range, and we confirmed nonclassical photon statistics for $\lambda_t =1300$\,nm. 
The customizable features of this PCF platform, i.e., fiber length, pump pulse wavelengths, bandwidths and energies, can all be set to optimize the conversion efficiency across a broad range of spectro-temporal scenarios. 
As such, we expect the PCF to coherently link many components in a hybrid quantum network, or assist quantum sensing at wavelengths with limited detector options.
Moreover, the demonstrated tuning range in this experiment covers over 150 independent frequency bins of the converted light.
This PCF is, therefore, also a promising candidate for broadband quantum optical processing in the frequency domain.

\begin{acknowledgments}
The authors are grateful for discussions with Khabat Heshami, Aaron Goldberg, Fr\'{e}d\'{e}ric Bouchard, Kate Fenwick, Guillaume Thekkadath, Yingwen Zhang, Denis Guay, Rune Lausten, Doug Moffatt, and Alex Davis. 
This work was supported by by the UK Hub in Quantum Computing and Simulation, part of the UK National Quantum Technologies Programme with funding from UKRI EPSRC grant EP/T001062/1.
%The authors acknowledge that this work was performed on the traditional unceded territory of the Algonquin Anishinaabe people.
\end{acknowledgments}

\bibliography{_QFCinPCF_PRLsubmission_20220727.bbl}

\end{document}